\documentclass{webofc}
\usepackage[varg]{txfonts}   

\usepackage{bm}
\newcommand{\bra}[1]{\langle \, #1 \, |}
\newcommand{\ket}[1]{| \, #1 \, \rangle}

\newcommand{\kket}[1]{ \, #1 \, \rangle}

\begin{document}
\title{Role of the effective range in the weak-binding relation}
%
%

\author{\firstname{Tomona} \lastname{Kinugawa}\inst{1}\fnsep\thanks{\email{kinugawa-tomona@ed.tmu.ac.jp}} \and
        \firstname{Tetsuo} \lastname{Hyodo}\inst{1}\fnsep\thanks{\email{hyodo@tmu.ac.jp}} 
}

\institute{Department of Physics, Tokyo Metropolitan University, Hachioji 192-0397, Japan
          }

\abstract{%
We study the range correction in the weak-binding relation, which relates the internal structure of hadrons with the scattering length and the binding energy. Utilizing the effective field theories, we show that the effective range originates from the derivative coupling interaction as well as from the channel coupling to the bare state, and that the different contributions are not distinguishable. By examining the compositeness in the effective field theories, it is demonstrated that the effective range induces the finite range correction for the weak-binding relation in addition to the previously known contributions. We thus propose to include the range correction in the uncertainty terms of the weak-binding relation.
}
\maketitle
%
\section{Introduction}
\label{intro}

The number of the observed hadrons has been continuously increasing~\cite{ParticleDataGroup:2020ssz}, and there are several candidates which seem to require exotic configurations beyond the simple $\bar{q}q$ or $qqq$ structures, such as multiquarks, hadronic molecules, hybrid states, and so on~\cite{Hosaka:2016pey,Guo:2017jvc}. The investigation of the exotic structures of hadrons is thus one of the major subjects in hadron physics. In particular, the recent observation of the doubly charmed tetraquark $T_{cc}$ by the LHCb collaboration~\cite{LHCb:2021vvq,LHCb:2021auc} further stimulates this trend.

Although there are many approaches to unveil the internal structure of hadrons, it is desirable to relate the hadron structure with observable quantities in a model-independent manner. One promising method is the weak-binding relation for bound states by Weinberg~\cite{Weinberg:1965zz}
\begin{align}
   a_{0}
   &=
   R\left(
   \frac{2X}{1+X}+\mathcal{O}\left(\frac{R_{\rm typ}}{R}\right)\right)
   \label{eq:WB} ,
\end{align}
where the compositeness $X$, the fraction of the hadronic molecule component, is related with the scattering length $a_{0}$ and the length scale $R=1/\sqrt{2\mu B}$ determined by the binding energy $B$. The higher order correction terms are of the order of $\mathcal{O}(R_{\rm typ}/R)$ with the typical length scale of the microscopic two-body interactions $R_{\rm typ}$. When the binding energy $B$ is sufficiently small such that $R_{\rm typ}/R\ll 1$, then the compositeness $X$ can be estimated by the observables, $a_{0}$ and $B$. This means that the internal structure of a bound state can be studied, even without having detailed knowledge of the wavefunction of the state. This feature should be useful in hadron physics, because the experimental data allow us to determine only a few scattering parameters at best. Indeed, the notion of the compositeness as well as the generalization of the weak-binding relation have been recently discussed in the context of hadron physics~\cite{Baru:2003qq,Hyodo:2011qc,Aceti:2012dd,Sekihara:2014kya,Kamiya:2015aea,Guo:2015daa,Kamiya:2016oao}.

The weak-binding relation is closely related to the low-energy universality for the systems with a large scattering length~\cite{Braaten:2004rn,Naidon:2016dpf}. It is known that, when the system has a shallow bound state, the scattering length can be much larger than the interaction scale, and the microscopic details of the system becomes irrelevant for the low-energy physics. It follows in the strict weak-binding limit $(B\to 0)$ that any lengths in the system should scale with the scattering length, which implies that $a_{0}=R$. The weak-binding relation \eqref{eq:WB} tells us how the ideal relation $a_{0}=R$ is modified by the coupling to the other components ($2X/(1+X)$ with $X< 1$) and by the finite range effect ($\mathcal{O}(R_{\rm typ}/R)$). In the previous studies, the length scale $R_{\rm typ}$ was estimated by the interaction range, which reflects the finite range correction from the off-shell nature of the system. On the other hand, there exists yet another length scale, the effective range, which reflects the range correction in the on-shell scattering amplitude. 

Here we would like to focus on the finite range correction in the weak-binding relation~\eqref{eq:WB}, stemming from the effective range~\cite{Kinugawa:2021ybb}. To this end, we examine the origin of the effective range in several effective field theories. With the analysis of the compositeness of the bound state, we then propose a method to include the range correction in the weak-binding relation.

\section{Effective field theories}
\label{sec:EFT}

Effective field theories are useful tools to describe the low-energy behaviors of given physical systems. We summarize properties of the nonrelativistic effective field theories for the low-energy two-body scattering following Ref.~\cite{Braaten:2007nq}, and show that the effective range can be induced either by the derivative coupling or the inclusion of the additional bare field.

\subsection{Zero-range model}
\label{sec:ZR}

The simplest effective field theory contains the four-point contact interaction, called the zero-range model, whose Hamiltonian is given by
\begin{align}
   \mathcal{H}_{\rm ZR}
   &=
   \frac{1}{2m}\bm{\nabla}\psi^{\dag}
   \cdot \bm{\nabla}\psi
   +\frac{\lambda_{0}}{4}
   (\psi^{\dag}\psi)^{2} ,
   \label{eq:zerorangeH} 
\end{align}
where $\psi$ is the boson field with mass $m$, and $\lambda_{0}$ is the bare coupling constant. The four-point contact interaction corresponds to the delta function potential in the coordinate space. Because of the singularity at the origin which is stronger than the kinetic term, the delta function potential in the three-dimensional space should be regularized to avoid the ultraviolet divergence. In the effective field theory, the momentum integration should be cut off at some ultraviolet scale $\Lambda$. By regarding Eq.~\eqref{eq:zerorangeH} as an effective description of the underlying microscopic theory with a finite range interaction,  we can interpret $\Lambda$ as the momentum scale below which the interaction can be well approximated by the pointlike one, as in Eq.~\eqref{eq:zerorangeH}. In other words, the inherent interaction range in the microscopic theory should be estimated by $\sim 1/\Lambda$. 

It is known that the field theory \eqref{eq:zerorangeH} is renormalizable in the two-body sector. Namely, one can absorb the cutoff $\Lambda$ dependence by the bare parameter $\lambda_{0}$, without modifying the observables. By taking the $\Lambda\to \infty$ limit after the renormalization, one finds that the two-body scattering amplitude in the zero-range model is determined exclusively by the scattering length $a_{0}$:
\begin{align}
   f_{\rm ZR}(k)
   &=
   \left(-\frac{1}{a_{0}}-ik\right)^{-1}.
   \label{eq:zerorangef} 
\end{align}
Because we interpret $1/\Lambda$ as the interaction range, the $\Lambda\to \infty$ limit corresponds to the zero range limit where the interaction range is formally set to zero. This model captures the basic aspects of the low-energy universality. In fact, Eq.~\eqref{eq:zerorangef} has a pole at $k=i/a_{0}$, which represents the bound state for a positive $a_{0}$. Noting that $R=i/k$, one finds the relation
\begin{align}
   a_{0}
   &=
   R
   \label{eq:zerorangeaR} ,
\end{align}
as indicated by the low-energy universality.

\subsection{Effective range model}
\label{sec:ER}

The first correction to the zero-range model can be obtained by introducing a derivative coupling term. This is called the effective range model:
\begin{align}
   \mathcal{H}_{\rm ER}
   &=
   \frac{1}{2m}\bm{\nabla}\psi^{\dag}
   \cdot \bm{\nabla}\psi
   +\frac{\lambda_{0}}{4}
   (\psi^{\dag}\psi)^{2}
   +\frac{\rho_{0}}{4}\bm{\nabla}(\psi^{\dag}\psi)
   \cdot\bm{\nabla}(\psi^{\dag}\psi) ,
   \label{eq:effectiverangeH} 
\end{align}
where $\rho_{0}$ is the coupling constant in the derivative coupling interaction. Performing the same renormalization procedure with the zero-range model, we end up with the scattering amplitude in the $\Lambda\to \infty $ limit:
\begin{align}
   f_{\rm ER}(k)
   &=
   \left(-\frac{1}{a_{0}}+\frac{r_{e}}{2}k^{2}-ik\right)^{-1}
   \label{eq:effectiverangef} ,
\end{align}
which now contains the effective range $r_{e}$, because the Hamiltonian~\eqref{eq:effectiverangeH} consists of two coupling constants $\lambda_{0}$ and $\rho_{0}$, and there is a sensible $\Lambda\to \infty$ limit where $a_{0}$ and $r_{e}$ remain finite. Note however that the interaction range is zero in the $\Lambda\to 0$ limit, and therefore, the effective range $r_{e}$ is induced by the derivative coupling term. We also note that the effective range in Eq.~\eqref{eq:effectiverangef} is always negative. This is consistent with the Wigner causality bound~\cite{Bohm,Matuschek:2020gqe}, which imposes the 
upper bound of the effective range to be of the order of the interaction range.

Because the denominator of Eq.~\eqref{eq:effectiverangef} is quadratic in $k$, the amplitude always contains two poles. Substituting $R=i/k$ for the bound state condition $1/f_{\rm ER}(k)=0$ and picking up the solution which is closer to the $k=0$, we obtain the relation between $a_{0}$, $R$, and $r_{e}$ as
\begin{align}
    a_0
    =R\frac{2r_e/R}{1-(r_e/R-1)^2}
    =R\frac{2}{2-r_e/R} .
    \label{eq:effectiverangea0}
\end{align}
This reduces to $a_{0}=R$ for the vanishing effective range $r_{e}\to 0$. For finite $r_{e}$, however, $a_{0}$ deviates from $R$. Because $R>0$, the negative effective range implies $a_{0}<R$.

\subsection{Resonance model}
\label{sec:R}

It is also possible to introduce an additional bare field $\phi$ which couples to the two-body scattering state,
\begin{align}
   \mathcal{H}_{\rm R}
   &=
   \frac{1}{2m}\bm{\nabla}\psi^{\dag}
   \cdot \bm{\nabla}\psi
   +\frac{1}{4m}\bm{\nabla}\phi^{\dag}
   \cdot \bm{\nabla}\phi
   +\nu_{0}\phi^{\dag}\phi
   +\frac{g_{0}}{2}(\phi^{\dag}\psi^{2}+\psi^{\dag 2}\phi)
   +\frac{\lambda_{0}}{4}
   (\psi^{\dag}\psi)^{2}
   ,
   \label{eq:resonanceH} 
\end{align}
where $\nu_{0}$ is the energy of the bare state measured from the threshold of the scattering state and $g_{0}$ is the coupling strength between the scattering state and the bare field. The scattering amplitude in the zero range limit $\Lambda\to \infty$ is obtained as
\begin{align}
   f_{\rm R}(k)
   &=
   \left(-\frac{1}{a_{0}}+\frac{r_{e}}{2}k^{2}+\mathcal{O}(k^{4})-ik\right)^{-1}
   ,
   \label{eq:resonancef} 
\end{align}
where $r_{e}$ and the higher order terms in the effective range expansion also appear in the amplitude, because the elimination of the bare field $\phi$ induces the energy dependence in the effective two-body interaction. Again in this case, the interaction range is set to zero, so the effective range originates in the coupling to the bare field. Also, the effective range is always negative, in accordance with the Wigner bound. To describe the system with a positive effective range, one needs to introduce the ghost field with negative norm
\cite{Kaplan:1996nv,Braaten:2007nq}.

In this model, the expression of $R$ cannot be written down in a closed form by the observables, because of the higher order terms in the effective range expansion. It is however clear that the $a_{0}=R$ relation does not hold in general. The expression of $R$ reduces to Eq.~\eqref{eq:effectiverangea0} when the higher order terms are neglected. 

\section{Weak-binding relation and range correction}
\label{sec:rangecorrection}

Now we consider the structure of the bound state in the above field theories from the viewpoint of the weak-binding relation and the compositeness. 

\subsection{Compositeness and the weak-binding relation}
\label{sec:WB}

As shown in Refs.~\cite{Kamiya:2015aea,Kamiya:2016oao} the Weinberg's weak-binding relation can be derived in the effective field theories. The compositeness $X$ is defined as the overlap integral of the bound state $\ket{B}$, which is the eigenstate of the full Hamiltonian, and the two-body scattering states $\ket{\bm{p}}$, the eigenstates of the free Hamiltonian:
\begin{align}
   X
   &=
   \int \frac{d^{3}\bm{p}}{(2\pi)^{3}}
   |\bra{\bm{p}}\kket{B}|^{2} ,
\end{align}
which is the probability of finding the two-body components in the bound state wavefunction.

We first consider the value of $X$ from the completeness relation in the effective field theories. It can be shown by the phase symmetry of the $\psi$ field in the Hamiltonians~\eqref{eq:zerorangeH} and \eqref{eq:effectiverangeH} that the particle number is conserved. This means that the completeness relation spanned by the eigenstates of the free Hamiltonian in the zero-range model and the effective range model for the two-body sector is given only by the scattering states:
\begin{align}
   1_{\rm ZR}
   &=1_{\rm ER}
   =\int \frac{d^{3}\bm{p}}{(2\pi)^{3}}
   \ket{\bm{p}}\bra{\bm{p}} .
\end{align}
Using the normalization condition for the bound state $\bra{B}\kket{B}=1$, we obtain the value of the compositeness 
\begin{align}
   X_{\rm ZR}
   &=X_{\rm ER}
   =1.
   \label{eq:XexactZRER}
\end{align}
In other words, by definition, the exact value of the compositeness is unity in the zero-range and effective range models. On the other hand, from the phase symmetry, the completeness relation in the resonance model is given by the sum of the bare state $\phi$ and the scattering states~\cite{Kamiya:2015aea,Kamiya:2016oao}:
\begin{align}
   1_{\rm R}
   &=
   \ket{\phi}\bra{\phi}
   +\int \frac{d^{3}\bm{p}}{(2\pi)^{3}}
   \ket{\bm{p}}\bra{\bm{p}} .
\end{align}
In this case, we obtain
\begin{align}
   X_{\rm R}
   &<
   1, 
   \label{eq:XexactR}
\end{align}
because $\bra{B}\kket{\phi}\bra{\phi}\kket{B}=|\bra{\phi}\kket{B}|^{2}>0$ provided that $g_{0}\neq 0$.

Next, we use the weak-binding relation~\eqref{eq:WB} to determine the compositeness $X$ in these theories. Recalling that the interaction range is related to the cutoff as $R_{\rm typ}=1/\Lambda$, in the zero range limit $\Lambda\to \infty$, the correction terms $\mathcal{O}(R_{\rm typ}/R)$ vanish, and Eq.~\eqref{eq:WB} determines $X$ only by $a_{0}$ and $R$. For $a_{0}=R$ in the zero-range model, we obtain $X_{\rm ZR}=1$, which coincides with the exact value in Eq.~\eqref{eq:XexactZRER}. On the other hand, in the effective range model, we obtain $a_{0}<R$, which corresponds to $X_{\rm ER}<1$. This is in contradiction to the exact value~\eqref{eq:XexactZRER}. For the resonance model, the weak-binding relation gives $X_{\rm R}<1$, in accordance with the exact value~\eqref{eq:XexactR}. 

The above results are summarized in Table~\ref{tbl:summary}. It is clear that the weak-binding relation~\eqref{eq:WB} with $R_{\rm typ}=1/\Lambda$ works for the zero-range model and the resonance model, but in the effective range model, the weak-binding relation does not reproduce the exact value. The discrepancy cannot be attributed to the correction terms which vanish in the zero range limit. The origin of the discrepancy is traced back to the finite effective range which violates the universal relation $a_{0}=R$. At the same time, in the resonance model, the finite effective range does not provide contradiction in the weak-binding relation, because the exact value of the compositeness is modified by the bare state contribution.

\begin{table}
\centering
\caption{Summary of the effective range $r_{e}$, the exact value of the compositeness $X_{\rm exact}$, and $X_{\rm WB}$ determined by the weak-binding relation~\eqref{eq:WB} in the effective field theories in the zero range limit. }
\label{tbl:summary}       
\begin{tabular}{llll}
\hline
EFT & $r_{e}$ & $X_{\rm exact}$ & $X_{\rm WB}$  \\ \hline
Zero-range model & $r_{e}=0$ & $X=1$ & $X=1$  \\ 
Effective range model & $r_{e}<0$ & $X=1$ & $X<1$\\
Resonance model & $r_{e}<0$ & $X<1$ & $X<1$ \\ \hline
\end{tabular}
\end{table}

\subsection{Origin of the effective range}
\label{sec:origin}

We have shown that the effective range $r_{e}$ originates either from the derivative coupling interaction (effective range model) or from the channel coupling to the bare state (resonance model). While the former does not modify the value of the compositeness, the latter induces $X_{\rm R}<1$ because of the bare state contribution. In other words, the origin of the effective range is two-fold. In order to determine the compositeness from the observables, one may wonder if we can distinguish the origin of the effective range. 

For this purpose, we consider the effective field theory which contains both the derivative coupling interaction in Eq.~\eqref{eq:effectiverangeH} and the coupling to the bare field in Eq.~\eqref{eq:resonanceH}. Calculating the effective range with a finite cutoff $\Lambda$, we obtain
\begin{align}
   r_{e}
   =\frac{16\pi}{m}
   \frac{\left[1+\frac{m}{12\pi^{2}}\rho_{0}\Lambda^{3}\right]^{2}
   \left\{2\rho_{0}\left[1+\frac{m}{24\pi^{2}}\rho_{0}\Lambda^{3}\right]
   -\frac{g_{0}^{2}}{m\nu_{0}^{2}}\right\}}
   {\left[
   \lambda_{0}-\frac{m}{20\pi^{2}}\rho_{0}^{2}\Lambda^{5}
   -\frac{g_{0}^{2}}{\nu_{0}}
   \right]^{2}} .
   \label{eq:range} 
\end{align}
In this expression, the derivative coupling contribution is in the terms with $\rho_{0}$, and the bare state contribution is in the $g_{0}^{2}$ term. By setting $\rho_{0}=0$ ($g_{0}= 0$), Eq.~\eqref{eq:range} reduces to the expression of the effective range (resonance) model. Because the terms with $\rho_{0}$ and those with $g_{0}^{2}$ are entangled in Eq.~\eqref{eq:range}, we conclude that the origin of the effective range is not distinguishable. 

\subsection{Range correction}
\label{sec:correction}

The above conclusion indicates that the discrepancy between $X_{\rm exact}$ and $X_{\rm WB}$ in the effective range model cannot be resolved by the correction of the $2X/(1+X)$ term with $r_{e}$, because its origin is not determined. Rather, the range correction should be included in the correction term $\mathcal{O}(R_{\rm typ}/R)$, by modifying the definition of $R_{\rm typ}$. Namely, there exists a new correction term, which does not vanish in the zero range limit, $\Lambda\to \infty$.

In the weak-binding situation with a large $R$, the scattering length in the effective range model~\eqref{eq:effectiverangea0} can be given by
\begin{align}
    a_0
    =R\left\{1+\mathcal{O}\left(\left|\frac{r_e}{R}\right|\right)\right\}.
\end{align}
This suggests to redefine the length scale $R_{\rm typ}$ as 
\begin{align}
   R_{\rm typ}&=\max\{R_{\rm int},R_{\rm eff}\},
   \label{eq:newRtyp}
\end{align}
with $R_{\rm int}=1/\Lambda$, the range of the interaction in the microscopic theory (off-shell length scale). In Eq.~\eqref{eq:newRtyp}, we propose to add $R_{\rm eff}$, which is the length scale in the effective range expansion, such as $|r_{e}|$ or the properly normalized coefficients of the higher order terms (on-shell length scale). With this range correction, the discrepancy in the effective range model is resolved, because there remain the correction terms $\mathcal{O}(R_{\rm eff}/R)$ even in the zero range limit $\Lambda\to \infty$ ($R_{\rm int}\to 0$). 

The new definition~\eqref{eq:newRtyp} reduces to the previous one when $R_{\rm eff}<R_{\rm int}$. In usual cases without fine tuning, the magnitude of the effective range is of the same order with the interaction range $R_{\rm int}$, where the $R_{\rm eff}$ term in Eq.~\eqref{eq:newRtyp} would not have a substantial effect. However, because there is no lower limit on the value of $r_{e}$, its magnitude can in principle be much larger than the interaction range such that $R_{\rm eff}\gg R_{\rm int}$. In fact, it is numerically demonstrated that the improved weak-binding relation with Eq.~\eqref{eq:newRtyp} has larger applicability than the previous one~\cite{Kinugawa:2021ybb}. The effective range model in the zero range limit is one of such examples. Thus, the improvement in Eq.~\eqref{eq:newRtyp} will be important for the systems with a sizable magnitude of the effective range.

\section{Summary}
\label{sec:summary}

We have discussed the role of the effective range in the weak-binding relation. The properties of the low-energy effective field theories are reviewed, with the emphasis on the effective range and its origin. It is shown that the effective range induced by the derivative coupling interaction does not affect the compositeness. This causes the discrepancy between the exact value of the compositeness and that given by the weak-binding relation in the effective range model in the zero range limit. We then propose the range correction of the weak-binding relation by modifying the correction terms, such that the above discrepancy is resolved, and the applicability of the weak-binding relation is enlarged. 

The improved relation can be applied to various hadron systems. For instance, the analysis of the $T_{cc}$ by the LHCb collaboration~\cite{LHCb:2021auc} reported the upper bound of the magnitude of the effective range to be $11$ fm. Also, a recent analysis of $X(3872)$ suggests that $r_{e}\sim -5$ fm~\cite{Esposito:2021vhu} (see, however, Ref.~\cite{Baru:2021ldu} on the uncertainty of the determination of the effective range in these analysis). These results imply the larger magnitude of the effective range than the hadron interaction scale of $1$ fm and the range correction of the weak-binding relation should be important. These systems are currently under investigation, and the results will be reported elsewhere in future.

%
%

%
%
%

\end{document}